\documentclass[aps,pra,twocolumn,showpacs]{revtex4-1}
\usepackage{xcolor, graphicx}
\usepackage{amsmath, amssymb}
\usepackage{ulem}
\usepackage[colorlinks=true,urlcolor=blue,citecolor=blue,linkcolor=blue]{hyperref}

\begin{document}

\title{Multitime correlation functions in nonclassical stochastic processes}

\author{F. Krumm}\email{fabian.krumm@uni-rostock.de}\affiliation{Arbeitsgruppe Theoretische Quantenoptik, Institut f\"ur Physik, Universit\"at Rostock, D-18051 Rostock, Germany}
\author{J. Sperling}\affiliation{Arbeitsgruppe Theoretische Quantenoptik, Institut f\"ur Physik, Universit\"at Rostock, D-18051 Rostock, Germany}
\author{W. Vogel}\affiliation{Arbeitsgruppe Theoretische Quantenoptik, Institut f\"ur Physik, Universit\"at Rostock, D-18051 Rostock, Germany}

\begin{abstract}
	A general method is introduced for verifying multitime quantum correlations through the characteristic function of the time-dependent $P$~functional that generalizes the Glauber-Sudarshan $P$~function.
	Quantum correlation criteria are derived which identify quantum effects for an arbitrary number of points in time.
	The Magnus expansion is used to visualize the impact of the required time ordering, which becomes crucial in situations when the interaction problem is explicitly time dependent.
	We show that the latter affects the multi-time-characteristic function and, therefore, the temporal evolution of the nonclassicality.
	As an example, we apply our technique to an optical parametric process with a frequency mismatch.
	The resulting two-time-characteristic function yields full insight into the two-time quantum correlation properties of such a system.
\end{abstract}

\pacs{
	42.50.-p, 
	03.65.Wj, 
	42.50.Xa, 
	03.65.Db  
}
\date{\today}
\maketitle

\section{Introduction}
	The determination of nonclassical effects, such as squeezing~\cite{W83} or entanglement~\cite{EPR35,S35}, is a cumbersome, yet necessary task for employing these phenomena in future quantum technologies.
	However, most approaches are restricted to single-time quantum effects, albeit it is well known that there exist multitime correlations~\cite{APT14} and quantum effects.
	For instance, the fundamental photon antibunching experiment requires correlating two points in time~\cite{KDM77}.

	To uncover the dynamics of nonclassicality, the treatment of multitime nonclassicality is therefore a desirable aspect to be studied. 
	Based on the well-known Glauber-Sudarshan $P$ function~\cite{G63,S63}, an approach to infer space-time-dependent quantum correlations of radiation fields was introduced by defining a generalized $P$~functional~\cite{V08}.
	The resulting nonclassicality criteria yield a general insight into multitime quantum correlation functions beyond the photon antibunching effect.

	Other characterization techniques have been introduced by Leggett and Garg~\cite{LG85}.
	They prove that temporal correlations can violate Bell-like equalities based on assumptions of macroscopic realism and noninvasive measurements.
	Such a test was recently applied to multilevel systems~\cite{BE14}.
	Another approach is the consideration of bits of classical communication that are required to simulate a temporal correlation function of a multilevel system~\cite{BKMPP15}.
	Considering other types of correlation functions, one can study the temporal dynamics of a coupled optomechanical system~\cite{SM16}.

	To keep the close relation between the treatment of the $P$~function and the $P$~functional, we will consider its Fourier transform, i.e., its  characteristic function.
	For a single point in time, it is known that the characteristic function yields necessary and sufficient nonclassicality certifiers~\cite{RV02}, which can be directly applied to measurements~\cite{LS02,MKMPE11}.
	Moreover, the characteristic function method can be related to moment-based techniques~\cite{SRV05} to formulate unified nonclassicality criteria~\cite{R15}.

	In the present contribution, we generalize the nonclassicality criteria in terms of characteristic functions to identify multitime nonclassical correlations.
	The derived conditions lead to an infinite hierarchy of necessary and sufficient nonclassicality probes for temporal quantum correlations in radiation fields.
	In general, the investigation of temporal correlations requires appropriate techniques, in particular, when the interaction problem under study is explicitly time dependent.
	For this purpose, we will apply the Magnus expansion of the unitary time-evolution operator to analyze the impact of time ordering on the evolution of nonclassicality.
	As a fundamental example, we provide a detailed study of the quantum correlations of a parametric process with a frequency mismatch.
	
	The paper is structured as follows.
	In Sec.~\ref{Sec:MTNCL} we recapitulate the concept of multitime nonclassicality.
	Necessary and sufficient conditions for time-dependent quantum correlations are derived in terms of multi-time-characteristic functions in Sec.~\ref{Sec:CFQC}.
	Section~\ref{Sec:PP} deals with the application of the methods to a parametric process with frequency mismatch.
	The corresponding time-dependent quantum effects are uncovered in Sec.~\ref{Sec:TDQE}.
	A summary and some conclusions are given in Sec.~\ref{Sec:Conclusions}.

\section{Multitime nonclassicality}
\label{Sec:MTNCL}
	A well-established definition of nonclassicality of a radiation field is related to the Glauber-Sudarshan $P$~function~\cite{G63,S63,TG65,M86,VW06},
	\begin{equation}\label{Eq:Pfunction}
		P(\alpha)=\langle \mathord{:} \hat \delta (\hat a -  \alpha) \mathord{:} \rangle.
	\end{equation}
        Herein, $\hat \delta$ denotes the operator-valued $\delta$ distribution, and $\hat a$ is the bosonic annihilation operator of the radiation mode under study. 
        The prescription $:\dots:$ orders creation operators to the left of annihilation operators. 
        On this basis, a quantum state is referred to as nonclassical if the $P$ function fails to have the properties of a classical probability density~\cite{TG65,M86}.
	That is, $P( \alpha)$ cannot be described by the classical theory of light.
	Note that a generalization to multimode fields is straightforward.

	In correlation measurements, observables become relevant which depend on a set $\{t_i\}_{i=1}^k$ of points in time.
	For example, two-time intensity correlation functions are recorded in photon antibunching experiments~\cite{KDM77}.
	Higher-order correlations of general field operators describe the full quantum statistics of the radiation field in chosen space-time points~\cite{V08,SH06}.
	Thus, to completely cover the multitime nonclassicality, we consider the $P$~functional~\cite{V08},
        \begin{equation}
        \label{Eq:genPfunctional}
		P\big[\{ \alpha(t_i) ; t_i\}_{i=1}^k\big]=\Big\langle  \begin{smallmatrix}
		\circ \\ \circ
		\end{smallmatrix}  \prod_{i=1}^k\hat \delta (\hat a(t_i) -  \alpha(t_i)) \begin{smallmatrix}
		\circ \\ \circ
		\end{smallmatrix} \Big\rangle . 
	\end{equation}
	The symbol $\begin{smallmatrix}\circ \\ \circ\end{smallmatrix} \dots \begin{smallmatrix}\circ \\ \circ\end{smallmatrix}$ represents the normal- and time-ordering prescriptions.
	Due to the latter, creation (annihilation) operators are sorted with increasing (decreasing) time arguments from left to right~\cite{VW06} as it occurs in the photocounting theory~\cite{KK64}.

	With the introduced functional, the normally and time-ordered expectation value of an arbitrary observable $\hat O=\hat O( \hat a (t_1),\dots, \hat a (t_k) )$, which depends on the bosonic creation and annihilation operators at different times, can be written as 
	\begin{eqnarray}
		\label{Eq:PProbDens}
		& & \langle  \begin{smallmatrix} \circ \\ \circ \end{smallmatrix} \hat O( \hat a (t_1),\dots, \hat a (t_k) ) \begin{smallmatrix} \circ \\ \circ \end{smallmatrix} \rangle
		\\ \nonumber
		&=&  \int d^2 \alpha_1 \dots \int d^2 \alpha_k O(\alpha_1,\dots, \alpha_k) P\big[\{ \alpha_i ; t_i\}_{i=1}^k\big],
	\end{eqnarray}
	where we identified $\alpha_i= \alpha(t_i)$.
	Using Eqs.~\eqref{Eq:genPfunctional} and~\eqref{Eq:PProbDens}, multitime nonclassicality can be defined as follows~\cite{V08}:
	A radiation field shows nonclassical correlation properties if and only if
	\begin{equation}
		\label{nclcondition}
			\exists \hat f: \langle \begin{smallmatrix}
		\circ \\ \circ
		\end{smallmatrix} \hat f^\dag \hat f \begin{smallmatrix}
		\circ \\ \circ
		\end{smallmatrix} \rangle <0 . 
	\end{equation}
	Here, $\hat f=\hat f( \hat a (t_1),\dots, \hat a (t_k) )$ connotes an operator function of the bosonic creation and annihilation operators at different times.
	Note that these nonclassicality conditions include the ones in Refs.~\cite{TG65,M86} for a single time.

\section{Characteristic functions and quantum correlations}
\label{Sec:CFQC}
\subsection{Space-time nonclassicality criteria}
	The Fourier transform of the multitime $P$~functional in Eq.~\eqref{Eq:genPfunctional} defines the corresponding characteristic function,
	\begin{equation}\label{Eq:CharFctDef}
		\Phi(\{ \beta_i;t_i\}_{i=1}^k)=\big\langle \begin{smallmatrix}
		\circ \\ \circ
		\end{smallmatrix} \hat D(\{ \beta_i;t_i\}_{i=1}^k) \begin{smallmatrix}
		\circ \\ \circ
		\end{smallmatrix} \big\rangle.
	\end{equation}
	Here, the multi-time-displacement operator reads as
	\begin{equation}\label{Eq:DisplOp}
		\hat D(\{ \beta_i;t_i\}_{i=1}^k)= \prod_{i=1}^k\exp\left[ \beta_i\hat a(t_i)^\dag -  \beta_i^\ast \hat a(t_i) \right]. 
	\end{equation}
	Characteristic functions are experimentally accessible, for example, via balanced homodyne correlation measurements~\cite{SH06} of the multi-time-characteristic function~\eqref{Eq:CharFctDef}.
	In the classical probability theory of stochastic processes, the coherent amplitudes $\{ \alpha_i(t_i) \}_{i=1}^k$ define a set of random variables, distributed according to the classical analog of the $P$~functional~\eqref{Eq:genPfunctional}.
	Their characteristic function~\eqref{Eq:CharFctDef} is a unique characterization in Fourier-transformed phase space, given by the variables $\{ \beta_i;t_i\}_{i=1}^k$.
	As the $P$ function at a single time can be highly singular, the same holds true for the general multitime functional.
	However, the characteristic function is always well behaved, which will also be discussed in the continuation of this paper and, therefore, much better suited for uncovering nonclassical features.

	We may expand each operator function in Eq.~\eqref{nclcondition} in terms of a Fourier series,
	\begin{equation}
		\hat f =  \sum_{j=1}^{ O} f_j \hat D(\{ \beta_{i,j};t_i\}_{i=1}^k) \label{Eq:Expansionf}
	\end{equation}
        for some complex amplitudes (index $j$) at different points in time (index $i$), $\beta_{i,j}$, and a given order $ O$.
	This expansion~\eqref{Eq:Expansionf} is the multitime generalization of the single-time case~\cite{SRV05,SV05}.
	Inserting Eq.~\eqref{Eq:Expansionf} into Eq.~\eqref{nclcondition} and using the definitions in Eqs.~\eqref{Eq:CharFctDef} and~\eqref{Eq:DisplOp}, we obtain
	\begin{eqnarray}
		\nonumber
		0>   \langle \begin{smallmatrix}
		\circ \\ \circ
		\end{smallmatrix} \hat f^\dag   \hat f \begin{smallmatrix}
		\circ \\ \circ
		\end{smallmatrix} \rangle
		&=&  \sum_{l,j=1}^{ O} f_l^\ast f_j \Phi(\{ \beta_{i,j}- \beta_{i,l};t_i\}_{i=1}^k) \\
		&=& \boldsymbol{f}^\dag{\boldsymbol{  \Phi}}(\{t_i\}_{i=1}^k) \boldsymbol{f},
		\label{matrixform}
	\end{eqnarray}
	where the latter vector-matrix notion consists of the matrix 
	${\boldsymbol{  \Phi}}(\{t_i\}_{i=1}^k)=[\Phi(\{ \beta_{i,j}- \beta_{i,l};t_i\}_{i=1}^k)]_{l,j=1}^{ O}$ and the coefficient vector $\boldsymbol{f}=(f_1,\dots,f_{ O})^{\rm T}$.

	Applying Bochner's theorem for classical probabilities~\cite{B33}, condition~\eqref{matrixform} certifies that the characteristic function~\eqref{Eq:CharFctDef} cannot be interpreted as the Fourier transform 
	of the classical analog of the $P$~functional~\eqref{Eq:genPfunctional}.
	Moreover, Bochner's conditions are necessary and sufficient if all orders $ O$ and all $\beta_{i,j}$ are considered.
	Applying Sylvester's criterion and in analogy to Ref.~\cite{RV02}, we get a multitime generalization of hierarchies of nonclassicality conditions.
	Namely, $\Phi(\{ \beta_i;t_i\}_{i=1}^k)$ is the characteristic function of a nonclassical $P$~functional if and only if
	\begin{equation}\label{Eq:mathcompact}
	    \exists \ {O} \in \mathbb{N}, \quad [\{\beta_{i,j}\}_{i=1}^{k}]_{j=1}^{O}: \det{}_{{O}} \left[{\boldsymbol{ \Phi}}(\{t_i\}_{i=1}^k)\right]<0.
	\end{equation}
	Herein, $\det{}_{O}$ denotes the determinant of the order $ O$.
	As $\det{}_1{[ {\boldsymbol{\Phi }}(\{t_i\}_{i=1}^k)]}=\Phi(\{0;t_i\}_{i=1}^k)=1$ (the normalization of $P$), the lowest-order nontrivial nonclassicality criterion is obtained for ${O}=2$. It reads
	\begin{equation} \label{Eq:lowestorder}
		0 > \det{}_2{\left[ {\boldsymbol{\Phi }}(\{t_i\}_{i=1}^k)\right]}=  1-  |\Phi( \{\beta_{i,2}-\beta_{i,1};t_i\}_{i=1}^k) |^2.
	\end{equation}
	This lowest-order criterion only provides a sufficient condition for nonclassicality, whereas the general nonclassicality condition~\eqref{Eq:mathcompact} is necessary and sufficient if all orders $ O$ are considered.
	Again, for one point in time, we recover the single-time nonclassicality condition established in Ref.~\cite{V00}.

	Let us relate the definition of nonclassicality of multiple points in time, i.e., the functional in Eq.~\eqref{Eq:genPfunctional} does not describe a classical stochastic process~\cite{V08} with the derived  criteria in inequality~\eqref{Eq:mathcompact}.
	The here introduced method is based on the characteristic function.
	In contrast to the $P$~functional, this function is always well behaved and does not exhibit singularities which are often present in the Glauber-Sudarshan representation.
	As the characteristic function includes all information about the quantum systems for the considered times, our approach can identify all nonclassical features which are present in the $P$~functional.
	In this sense, our method is equivalent to the approach in Ref.~\cite{V08}, which, however, formulates the conditions for nonclassical correlations in terms of time-dependent correlation functions.
	Alternatively, the here presented technique relies on a regular phase-space distribution in terms of the characteristic function and the corresponding criteria in~\eqref{Eq:mathcompact}.

\subsection{Time evolution and time ordering}
	In order to rigorously study the propagation of quantum correlations in time, let us formulate the dynamic behavior of quantum systems.
	The most general way to implement the propagation in time is formulated in terms of the unitary time-evolution operator $\hat {\mathcal{U}}(t)$.
	For an explicitly time-dependent Hamiltonian $\hat H(t)$, the latter is given by
	\begin{equation}\label{Eq:timeorderU}
		 \hat {\mathcal{U}}(t)= \mathcal{T}\exp\left[-\frac{i}{ \hbar}\int_0^t dt' \hat H(t')\right],
	\end{equation}
	where $\mathcal{T}$ denotes exclusively the time-ordering prescription, which is required if $[\hat H(t_1), \hat H(t_2) ] \neq 0$ for $t_1 \neq t_2$.
	As such a formal solution is not very helpful, one can apply the Dyson expansion, $\hat {\mathcal{U}}(t)=\hat 1-i\hbar^{-1}\int_0^t dt_1\hat H(t_1)-\hbar^{-2}\int_0^t dt_1\int_0^{t_1} dt_2\hat H(t_1)\hat H(t_2)+\dots$, 
	which resembles a time-ordered Taylor expansion of the exponential function in~\eqref{Eq:timeorderU}.
	Alternatively, this problem can be handled by the Magnus expansion~\cite{M54,B09}, which yields the time-evolution operator in the form
	\begin{equation}\label{Eq:Magnusgeneral}
		\hat {\mathcal{U}}(t)=\exp\left[\sum_{n=1}^{n_{\rm max}} \hat \Omega_n(t)\right],
	\end{equation}
	with $\hat \Omega_n(t)$ being the $n$th Magnus order and for $n_{\rm max}=\infty$, the true evolution is recovered.
	In contrast to the Dyson series, the Magnus expansion is unitary in each order.
	The first two orders read as
	\begin{eqnarray}
		\hat \Omega_1(t)&=&-\frac{i}{ \hbar} \int_0^t dt_1 \hat H(t_1), \nonumber \\
		\hat \Omega_2(t)&=&- \frac{1}{ 2 \hbar^2}  \int_0^t dt_1 \int_0^{t_1} dt_2 [\hat H(t_1), \hat H(t_2) ].
		\label{Eq:Magnus2ndOrderCorrection}
	\end{eqnarray}
	Hence, nonequal time commutators of the Hamiltonian $\hat H(t)$ have to be evaluated.
	The first-order approximation~\eqref{Eq:Magnusgeneral} $n_{\max}=1$ is equivalent to disregarding the time-ordering $\mathcal{T}$ in Eq.~\eqref{Eq:timeorderU}.
	Higher Magnus orders, $\hat \Omega_{n}(t)$ with $n>1$, can be referred to as \textit{time-ordering corrections}.

	Thus, an effect of the time ordering can be observed if the higher-order corrections are nonzero.
	In particular for the second order, this means that the Hamiltonians of the system for different times do not commute, see Eq.~\eqref{Eq:Magnus2ndOrderCorrection}.
	The Magnus expansion and the hierarchy of introduced nonclassicality criteria~\eqref{Eq:mathcompact} --including the special case~\eqref{Eq:lowestorder}-- enable us to directly investigate the impact of time-ordering effects on the nonclassicality.

\section{Application to parametric processes} 
\label{Sec:PP}
	As a fundamental example, we study the particular process of degenerate parametric down-conversion for a strong classical pump field in the following.
	In the interaction picture, the time-dependent interaction Hamiltonian reads as
        \begin{equation}\label{Eq:IntHamil}
		\hat H_{\rm int}(t)=\hbar \kappa  \left[ e^{-i\delta t}\hat a^{\dag 2} + e^{i\delta t}\hat a^{2} \right],
        \end{equation}
        where $ \kappa$ denotes the coupling constant and $\delta=\omega_{\rm p}-2\omega_{\rm a}$ is a nonlinear frequency mismatch between the pump frequency $\omega_{\rm p}$ and the signal frequency $\omega_{\rm a}$. 
        For a perfectly matched pump, $\delta=0$, Eq.~\eqref{Eq:timeorderU} without the time ordering already gives the exact solution.

        A related system has been studied recently~\cite{QS14,QS15}.
	There, a two-mode Hamiltonian was considered with the frequency spectrum using the Magnus expansion.
	In Ref.~\cite{CBMS13}, the authors have also studied parametric down-conversion with respect to time-ordering corrections.
	In contrast to those works, we aim at demonstrating the impact of time ordering on the continuous evolution of quantum phenomena.

	Equivalent to the time-evolution operator in Eq.~\eqref{Eq:timeorderU}, one can establish the coupled equation of motion in the interaction picture for the signal field operators $\hat a$ and $\hat a^\dag$. This reads as
	\begin{eqnarray}\nonumber
		\frac{d}{dt} \begin{pmatrix}
			\hat a(t) \\
			\hat a(t)^\dag
		\end{pmatrix}
		&=& \begin{pmatrix}
		0 & -2i \kappa e^{-i\delta t} \\
		2i \kappa e^{i\delta t}& 0
		\end{pmatrix}
		\begin{pmatrix}
			\hat a(t) \\
			\hat a(t)^\dag
		\end{pmatrix}
		\\\label{Eq:MatrixM}
		&=& \boldsymbol{M}(t)\begin{pmatrix}
			\hat a(t) \\
			\hat a(t)^\dag
		\end{pmatrix}.
	\end{eqnarray}
	Now, the formal solution in Eq.~\eqref{Eq:timeorderU} is given by $-i\hbar^{-1}\hat H(t) \mapsto  \boldsymbol{M}(t)$ which also replaces the operator $\hat {\mathcal{U}}(t)$ by the unitary matrix $\boldsymbol{\mathcal{U}}(t)$.
	This approach allows us to express the time evolution in terms of matrices which, then, can be directly applied to $\hat a$ and $\hat a^\dag$.
	In the same way $[\hat H(t_1),\hat H(t_2)]\neq0$ translates to non commuting matrices $ \boldsymbol{M}(t_1)$ and $ \boldsymbol{M}(t_2)$.

	For a rigorous formulation of our treatment, one has to mention the convergence radius of the Magnus series.
	The series converges if the inequality,
	\begin{equation}
	    \int_0^t ds \| \boldsymbol{M}(s) \|_2 <\pi
	\end{equation}
	is fulfilled~\cite{B09}.
	The convergence radius is the supreme value for which this inequality is satisfied.
	Using Eq.~\eqref{Eq:MatrixM}, one obtains the spectral norm as $\vert\vert \boldsymbol{M}(s) \vert\vert_2 = 2 \kappa$.
	For convenience we introduce the dimensionless time $\tau = 2 \kappa t/\pi$.
	Thus, convergence is assured for $\tau <1$.
	To access times $\tau \geq 1$, one can subdivide the evolution in adjacent time intervals, known as the Euler method.

	For our studied model, the Magnus expansion is completely determined by $\tau$ and $\delta/\kappa$.
	This means that the time evolution is given by the ratio of the mismatch and the coupling constant as well as the characteristic system time $\pi/(2\kappa)$.
	Using the matrix $\boldsymbol M(t)$ in Eq.~\eqref{Eq:MatrixM}, we explicitly show that each Magnus order has the form
	\begin{eqnarray}\label{Eq:OrdersForm}
		\boldsymbol\Omega_n(\tau)=\left\lbrace\begin{array}{ll}
			i|C_n|\begin{pmatrix}-1&0\\0&1\end{pmatrix} &\text{for $n$ even,}\\
			|S_n|\begin{pmatrix}0&e^{i\varphi_n}\\e^{-i\varphi_n}&0\end{pmatrix} &\text{for $n$ odd,}
		\end{array}\right.
	\end{eqnarray}
	where we use $\tau$ instead of the time $t$ for parametrization. The quantities $|C_n|$, $|S_n|$, and $\varphi_n$ are obtained numerically, for details see Ref.~\cite{B09}.

	With Eq.~\eqref{Eq:OrdersForm}, the time-evolution matrix reads as
	\begin{eqnarray}
		\nonumber\boldsymbol{\mathcal{U}}(\tau)&=& \exp\left[\sum_{n}\boldsymbol\Omega_n(\tau)\right]
		=\exp\left[\boldsymbol\Omega(\tau)\right]
		\\&=&\cosh(p)\begin{pmatrix}1&0\\0&1\end{pmatrix}
		+\frac{\sinh(p)}{p}\begin{pmatrix}-iC&S\\S^\ast&iC\end{pmatrix},
		\label{Eq:MatrixExp}
	\end{eqnarray}
	with $C=\sum_{n\text{ even}} |C_n|$, $S=\sum_{n\text{ odd}} |S_n|e^{i\varphi_n}$, and $p=[-\det \boldsymbol \Omega(\tau)]^{1/2}=[|S|^2-C^2]^{1/2}$.
	The nondiagonal entries of $\boldsymbol{\mathcal{U}}(\tau)$ correspond to a map $\hat a \mapsto \hat a^\dag$ (and $\hat a^\dag \mapsto \hat a$).
	Such a transformation yields squeezing.
	The diagonal elements can be related to rotations in phase space, which is a classical effect.
	
	An additional advantage of the dynamical representation in terms of Eq.~\eqref{Eq:MatrixM} and its Magnus expansion is the possibility of analyzing non-equal-time-commutation relations of the fundamental bosonic annihilation (or creation) operators,
	\begin{eqnarray}
		\nonumber
		& & [\hat a(\tau),\hat a(\tau+\Delta \tau)]
		\\&=&{\mathcal{U}}_{11}(\tau) {\mathcal{U}}_{12}(\tau+\Delta \tau)- {\mathcal{U}}_{12}(\tau) {\mathcal{U}}_{11}(\tau+\Delta \tau),
		\label{Eq:NonEqualTimeCommutation}
	\end{eqnarray}
	where the components of the evolution operator are used $\boldsymbol{\mathcal{U}}(\tau)=(\mathcal{U}_{ij}(\tau))_{i,j=1}^2$.
	The commutator~\eqref{Eq:NonEqualTimeCommutation} is, in general, nonzero for time-dependent Hamiltonians and unequal times $\Delta \tau \neq 0$.
	Such commutation relations for different points in time play a fundamental role in multitime correlation functions, see Ref.~\cite{VW06} for an overview.

	\begin{figure}[t]
	\centering
	\includegraphics*[width=8cm]{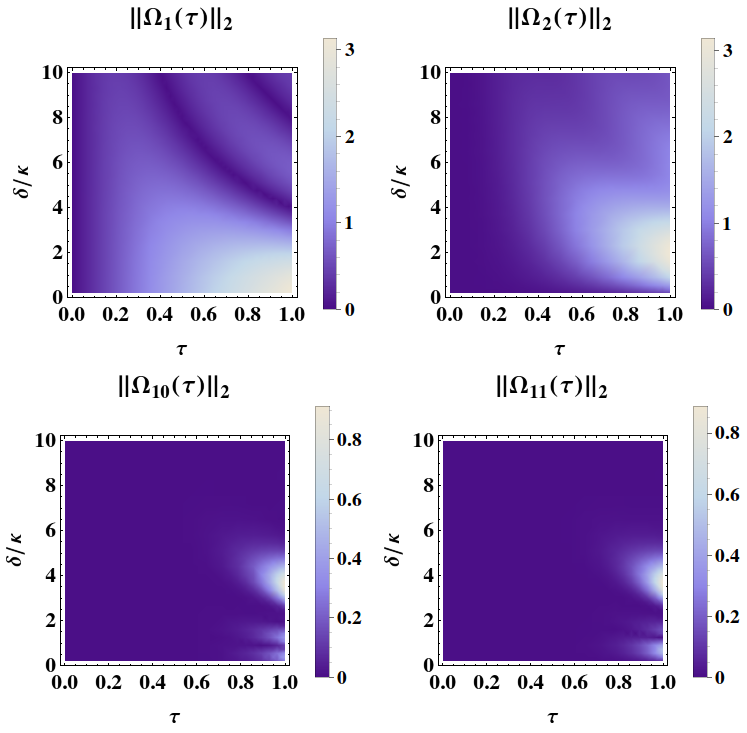}
	\caption{(Color online)
		Density plots of the spectral norm of different Magnus orders $\|\boldsymbol{\Omega}_n(\tau)\|_2$ for $n=1,2,10,11$.
		The time-ordering corrections are most significant at values of $\tau$ and $\delta/\kappa$, where $\|\boldsymbol{\Omega}_n(\tau)\|_2$ for $n>1$ attains its largest values.
	}\label{Fig:Magnusnorms}
        \end{figure}
        
	The impact of the different Magnus orders~\eqref{Eq:OrdersForm} is visualized in Fig.~\ref{Fig:Magnusnorms} by showing the spectral norm of different contributions $\|\boldsymbol{\Omega}_n(\tau)\|_2$.
	A large value of this matrix norm relates to a more significant contribution to the evolution operator~\eqref{Eq:MatrixExp}.
	The plot $n=1$ describes the case without any time ordering.
	The norms for $n>1$ are maximal in the range of $1 \leq \delta/\kappa \leq 5$.
	Thus, the strongest time-ordering corrections are expected to occur in this region.
	In particular, we will use the value $\delta/\kappa\approx 3.18$ for the following discussions.

\section{Time-dependent quantum effects}
\label{Sec:TDQE}
	After discussing the modeling of the evolution with time-dependent Hamiltonians, let us come back to the verification of nonclassicality.
	According to our criterion~\eqref{Eq:lowestorder}, we are able to verify nonclassical multitime correlations if $|\Phi( \{\beta_{i,2}-\beta_{i,1};t_i\}_{i=1}^k)|^2>1$.
	Let us apply this test to our considered system.
	Again, we will replace the times $t_i$ with the scaled ones $\tau_i=2 \kappa t_i/\pi$.

	First, we study the evolution of the nonclassicality by considering a single time $k=1$ for discussing the time-ordering corrections.
	Initially, the system is assumed to be in the vacuum state.
	Thus, after a principal axis transformation of $\beta=\beta_{1,2}-\beta_{1,1}$ to a rotated complex amplitude $\gamma$, we can write the modulus of the characteristic function~\eqref{Eq:CharFctDef} in the form
	\begin{equation}\label{eq:lambda_max}
		|\Phi(\gamma;\tau)|^2=e^{ \lambda ({\rm Re} \gamma)^2 + \mu ({\rm Im}\gamma)^2}.
	\end{equation} 
	If the maximal value $\lambda_{\rm max}=\lambda_{\rm max}(\tau)=\max (\lambda,\mu)$ is positive, then the characteristic function exceeds one.
	
\begin{figure}[t]
	\centering
	\includegraphics*[width=8cm]{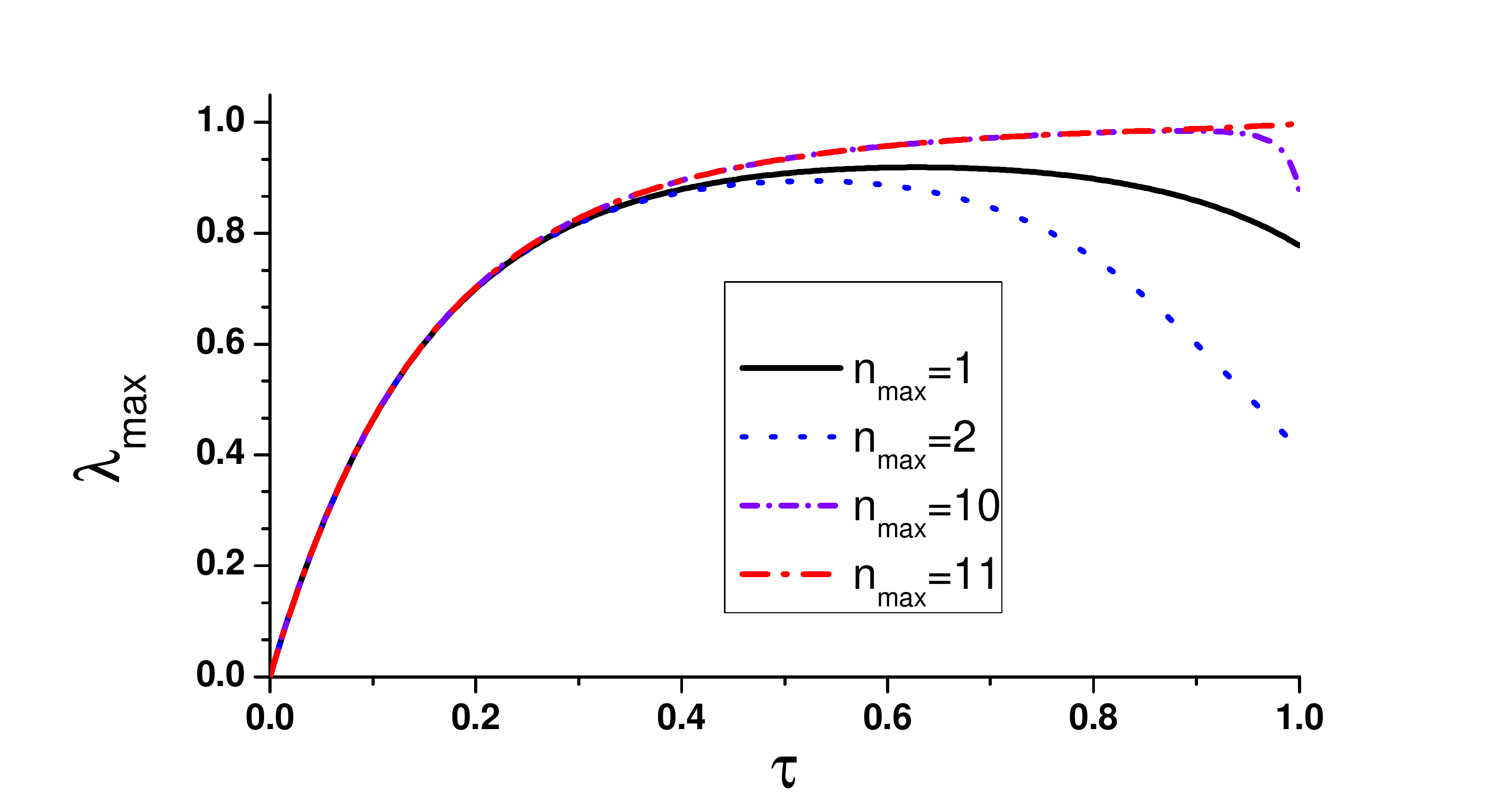}
	\caption{(Color online)
		$\lambda_{\rm max}$, cf. Eq.~\eqref{eq:lambda_max}, for $ \boldsymbol{\Omega}(\tau)=\sum_{n=1}^{n_{\rm max}}  \boldsymbol{\Omega}_n(\tau)$ and $\delta/\kappa \approx 3.18$ in dependence on the rescaled time $\tau$.
		A positive $\lambda_{\rm max}$ verifies nonclassicality.
		The influence of time ordering is clearly visible in the contributions of the higher-order Magnus terms $n_{\rm max}>1$.
	}\label{Fig:MaxEW}
\end{figure}

	The results for different Magnus orders are given in Fig.~\ref{Fig:MaxEW}, where $\lambda_{\rm max}$ is plotted as a function of $\tau$.
	One can see the time-ordering effects in the evolution for $n_{\rm max}>1$, which significantly affect the rising behavior of the characteristic function and, therefore, the nonclassical character of the corresponding state.
	Note that times close to the convergence radius $\tau=1$ show oscillations when comparing even and odd $n_{\rm max}$ values because of the alternating form of the contributions $\boldsymbol\Omega_n(\tau)$ in Eq.~\eqref{Eq:OrdersForm}.

	Second, we study two-time nonclassical correlations.
	We follow the same approach, such as in the single-time case, i.e., a principal axis transformation: $(\beta_{1,2}-\beta_{1,1},\beta_{2,2}-\beta_{2,1}) \mapsto (\gamma_1,\gamma_2)$.
	Consequently, the two-time characteristic function for the nonclassicality condition~\eqref{Eq:lowestorder} reduces to a form similar to~\eqref{eq:lambda_max},
	\begin{equation}\label{Eq:TwoTimeMS}
	    |\Phi( \{\gamma_i;\tau_i\}_{i=1}^2)|^2=\exp \left({\sum_{i=1}^2 [\lambda_i ({\rm Re} \gamma_i)^2 + \mu_i ({\rm Im} \gamma_i)^2]}\right) .
	\end{equation}

\begin{figure}[t]
	\centering
	\includegraphics*[width=8cm]{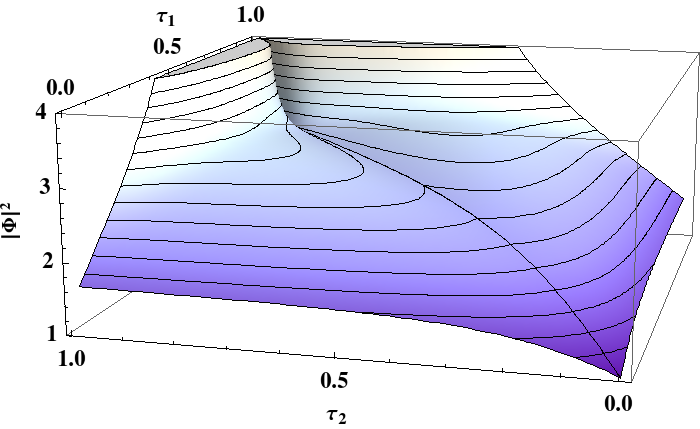}
	\caption{(Color online)
		We show $|\Phi|^2 \equiv |\Phi(\{\gamma_i;\tau_i\}_{i=1}^2)|^2$.
		As the value of one is exceeded for all times $\tau_{1,2} \neq 0$, the state is clearly two-time quantum correlated.
		We included all Magnus orders up to $n_{\rm max}=11$ in the evolution for $\delta/\kappa \approx 3.18$.
	}\label{Fig:MultitimeChar}
\end{figure} 
        The modulus square of the two-time characteristic function is shown in Fig.~\ref{Fig:MultitimeChar}.
        Note that due to the time-ordering prescription in Eq.~\eqref{Eq:CharFctDef}, one has to consider actually two scenarios $\tau_1 \geq \tau_2$ and $\tau_2 \geq \tau_1$.
        The quantity~\eqref{Eq:TwoTimeMS} depends on two complex parameters $(\gamma_1,\gamma_2)$.
        We restricted ourselves to a unit circle ($|\gamma_1|^2+|\gamma_2|^2=1$) and selected phase values ($\arg\gamma_1=\arg\gamma_2=\pi/2$), which maximize the function~\eqref{Eq:TwoTimeMS}.
	As the two-time-characteristic function is larger than one, we directly uncover two-time quantum correlations for all times, except for $\tau_{1,2}=0$.
	It is important to mention that the corresponding $P$~functional will be highly singular as an inverse Fourier transform of $\Phi$ is only possible in terms of singular distributions 
	as Eq.~\eqref{Eq:TwoTimeMS} is an unbounded function for $\max\{\lambda_1,\lambda_2,\mu_1,\mu_2\}>0$.
	Our results clearly visualize the two-time quantum correlations of our system.

	Finally, another advantage of our approaches is the fact that measurement schemes, e.g., as proposed in Ref.~\cite{SH06}, can be implemented which enables us to sample multi-time-characteristic functions of arbitrary orders.
	This allows one to directly implement our nonclassicality criteria~\eqref{Eq:mathcompact}.
	Thus, quantum correlations between an arbitrary number of points in time are experimentally accessible and reconstructible.

\section{Conclusions}
\label{Sec:Conclusions}
	To summarize, we presented techniques for identifying nonclassical stochastic processes in radiation fields in terms of characteristic functions.
	The latter are the Fourier transforms of the multi-time-dependent $P$~functional, generalizing the Glauber-Sudarshan representation.
	We constructed a hierarchy of necessary and sufficient criteria for the characteristic functions that can be applied to infer nonclassical correlations between an arbitrary number of points in time.
	The influence of the time ordering on the evolution of nonclassicality has been studied via the Magnus expansion.

	To demonstrate the applicability, we studied a degenerated parametric-down-conversion process with a frequency mismatch.
	We identified time intervals where the time-ordering corrections have the most significant impact for our time-dependent Hamiltonian.
	Whenever the temporal evolution of nonclassicality is studied in the presence of a Hamiltonian not commuting with itself at different points in time, this influences the nonclassicality of the system.
	Based on the characteristic function, we studied this temporal behavior of the nonclassicality and demonstrated nonclassical correlations between two points in time.

	Our method can be generalized in a straightforward way to simultaneously study multimode and multitime correlations.
	Moreover, the decomposition of the light field into a free-field and a source-field contribution allows one to study nonclassical processes in light-matter interactions.
	Especially the investigation of non-equal-time commutation relations -- including the description of a dynamic source -- is an interesting problem for future studies.
	Eventually, this will lead to a deeper understanding of the nonclassical evolution of quantum systems and their resulting nonclassical correlations.

\section*{Acknowledgments}
	This work has received funding from the Deutsche Forschungs\-gemeinschaft through SFB 652 and the European Union's Horizon 2020 research and innovation program under Grant Agreement No. 665148.

\end{document}